\documentclass[12pt]{article}

\begin{document}
\title{A possible explanation of a broad $1^{--}$ resonant structure around 1.5GeV}
\author{Bing An Li\\
Department of Physics and Astronomy, University of Kentucky\\
Lexington, KY 40506, USA}

\maketitle
\begin{abstract}
The broad $1^{--}$ resonant structure around 1.5GeV observed in the $K^+ K^-$ mass spectrum
in $J/\psi\rightarrow K^+ K^-\pi^0$ by BESII is interpreted as a composition of $\rho(1450)$
and $\rho(1700)$. Various tests are investigated.
\end{abstract}

\newpage

Recently, BESII has reported an observation of a broad resonant structure around 1.5GeV in 
the $K^+ K^-$ mass spectrum in $J/\psi\rightarrow K^+ K^-\pi^0$[1]. The quantum numbers of
this structure are determined to be $1^{--}$. In Ref.[1] single pole is used to fit the data
\begin{equation}
m=1576^{+49+98}_{-55-91} MeV\;\;\;\Gamma=818^{+22+64}_{-23-133}MeV.
\end{equation}
The branching ratio is determined to be
\begin{equation}
B(J/\psi\rightarrow X\pi^0)B(X\rightarrow K^+ K^-)=(8.5\pm0.6^{+2.7}_{-3.6})\times10^{-4}.
\end{equation}
Interpretation of this broad structure is a challenge.
In Ref.[2] this structure is interpreted as a $K^*(892)$-$\kappa$ molecule. Tetraqaurk state[3]
and diquark-antidiquark[4] are proposed to explain the broad width of this structure.
It is pointed out in Ref.[5] that two broad overlapped resonances $\rho(1450)$ and $\rho(1700)$ are right
at the region of the broad structure and have the same quantum numbers.
The final state interactions of $\rho(1450,1700)\rightarrow K^+ K^-$ are studied in Ref.[5] and 
the $B(J/\psi\rightarrow\pi^0\rho(1450,1700))B(\rho(1450,1700)\rightarrow K^+ K^-)\sim10^{-7}$ is obtained.
This branching ratio is far less than the experimental value(2).

In the range $\rho(1600)$ there is complicated structure.
A lot of strong evidences show that the 1600-MeV region actually contains two $\rho$-like resonances:
$\rho(1450)$ and $\rho(1700)$[6].
In this letter the possibility that the broad structure mentioned above is caused by 
$\rho(1450)$ and $\rho(1700)$ is revisited. The arguments are following:
\begin{enumerate}
\item Their quantum numbers are $1^{--}$ which are the same as the ones of the 
structure. They are isovectors and can be produced in $J/\psi\rightarrow\rho(1450,1700)\pi$.
\item Their masses are in the region of the structure.
\item The decay mode of $\rho(1450,1700)\rightarrow\bar{K}K$ has been found[7]. Therefore, these two
resonances do contribute to $J/\psi\rightarrow K\bar{K}\pi$.
\item Can the contributions of $\rho(1450, 1700)$ explain the 
$BR(J/\psi\rightarrow X\pi^0)B(X\rightarrow K^+ K^-)$(2)? 
This is a very important issue.
In Ref.[1] the estimation of $BR(J/\psi\rightarrow\rho(1450)\pi^0)BR(\rho(1450)\rightarrow K^+ K^-)
=(5.0\pm0.4)\times10^{-4}$ is obtained if the broad structure is from $\rho(1450)$.
This value is within the lower end of the value(2). The contribution of $\rho(1700)$  should be included.
$\rho(1450,1700)\rightarrow K^+ K^-$ are found in $\bar{p}p\rightarrow\rho(1450,1700)
\rightarrow K^+ K^-\pi^0$ in Ref.[8] and 
$B(\bar{p}p\rightarrow\rho(1700)\pi\rightarrow K^+ K^-\pi^0)=(2.9\pm0.8)\times10^{-4}$ is reported[8].
In Ref.[9] $BR(\bar{p}d\rightarrow\pi^-\pi^-\pi^+ p_{spectator})=(1.1\pm0.1)\times10^{-2}$ is presented.  
$\rho(770), \rho(1450), \rho(1700), f_2(1275)...$ are found in this process. The data in Ref.[7] leads to 
$BR(\rho(1700)\rightarrow\pi\pi)\sim0.25\times10^{-2}$. Using all these data, the estimation
$BR(\rho(1700)\rightarrow K\bar{K})>10^{-3}$ is obtained. Adding the contribution of $\rho(1700)$ to the 
estimation made in Ref.[1], the possibility that 
$BR(J/\psi\rightarrow\rho(1450,1700)\pi^0\rightarrow K^+ K^-\pi^0)$ reaches the value presented in Eq.(2) is not
impossible.

In Ref.[5] loop diagrams are calculated to determine the decay rates of $\rho(1450,1700)\rightarrow K\bar{K})$.
Very small branching ratios are found and the authors conclude that comparing with the data(2),
$BR(J/\psi\rightarrow\rho(1450,1700)\pi^0\rightarrow K^+ K^-\pi^0)$ is too small. A different point of view
is presented in this letter. $\rho(770)$ meson is made of u and d quarks.
In a chiral field theory of pseudoscalar, vector, and axial vector mesons[10] it is shown that
$\rho(770)$ is coupled to $K\bar{K}$ at the tree level[10]
\begin{eqnarray}
{\cal L}_{\rho K\bar{K}}&=&{2\over g}f_{\rho}(q^2)f_{iab}\rho^i_\mu K_a\partial_\mu K_b,\nonumber \\
{\cal L}_{\rho \pi\pi}&=&{2\over g}f_{\rho}(q^2)\epsilon_{ijk}\rho^i_\mu \pi^j\partial_\mu \pi^k,\nonumber \\
f_\rho(q^2)&=&1+\frac{q^2}{2\pi^2 m^2_\rho}\{(1-{2c\over g})^2-4\pi^2 c^2\},\nonumber \\
c&=&\frac{f^2}{2gm^2_\rho},
\end{eqnarray}
where g is a universal coupling constant and determined to be 0.39. 
Eqs.(3) show that in the chiral limit the strengths of the couplings
$\rho\pi\pi$ and $\rho K\bar{K}$ are the same at the tree level. In this chiral theory $\omega$ meson couples
to $K\bar{K}$ at the tree level too. All $\rho K\bar{K}$, $\omega K\bar{K}$, and $\phi K\bar{K}$ couplings 
contribute to both the form factors of charged kaon and neutral kaon[11]. Theory agrees with data very well.
The decay $\tau\rightarrow K\bar{K}\nu$ is dominated by the vertex ${\cal L}_{\rho K\bar{K}}$ 
at the tree level. Theory is in good agreement with data[12]. The vertex ${\cal L}_{\rho K\bar{K}}$
at the tree level contributes to $\pi K$ scatterings too[13] and good agreement with data is obtained[13].
Therefore, the coupling $\rho K\bar{K}$ at the tree level exists and is supported by experiments.
In this chiral field theory mason vertices at the tree level are at the leading order in $N_C$ expansion 
and loop diagrams of mesons are at higher order. Therefore, loop diagrams of mesons are suppressed by $N_C$ 
expansion. For example, $m_\omega-m_\rho$, $m_{f_1(1285)}-m_{a_1}$, $BR(\phi\rightarrow\rho\pi)$ are from
one-loop diagrams of mesons and they are small. Comparing with $\rho(770)$, $\rho(1450, 1700)$ are isovectors 
too. There is no obvious reason why
the couplings $\rho(1450,1700)K\bar{K}$ at the tree level are forbidden. On the other hand,  
based on the arguments of Ref.[10], loop diagrams of mesons are at higher order in $N_C$ expansion. Therefore,
$BR(\rho(1450,1700)\rightarrow K\bar{K})$ obtained from loop diagrams of mesons in Ref.[5] are small. 
Adding tree diagrams which at the leading order in $N_C$ expansion, 
the increase of $BR(\rho(1450,1700)\rightarrow K\bar{K})$ should be expected.
\item The experimental values of the widths of $\rho(1450,1700)$ have a wide range[7]. 
The range of the width of $\rho(1450)$ is $60-547\pm86^{+46}_{45}$MeV[7] and for $\rho(1700)$ it is 
$100-850\pm200$ MeV[7]. 
In a high statistics study of the decay $\tau^-\rightarrow\pi^-\pi^0\nu_\tau$[14] both $\rho(1450, 1700)$
are found. In one fit $\Gamma(\rho(1450))=471\pm29\pm21MeV$ and $\Gamma(\rho(1700))=
255\pm19\pm79MeV$ are determined.
$\Gamma(\rho(1450))=553\pm31\pm21MeV$ and $\Gamma(\rho(1700))=567\pm81\pm79MeV$ are obtained in the second fit.
Therefore, it is not a problem to use $\rho(1450,1700)$ to understand the broad structure(1).
\end{enumerate}
All the arguments provided above show it is possible that the broad structure observed in 
$J/\psi\rightarrow K^+ K^-\pi^0$ is caused by $\rho(1450,1700)$.  

In this letter tests of this possibility are investigated. The decay mode of 
$\rho(1450,1700)\rightarrow\pi\pi$ has been found[7]. Therefore, the broad structure 
observed in $J/\psi\rightarrow K^+K^-\pi^0$ should be observed
in the spectrum of $\pi^+\pi^-$ of $J/\psi\rightarrow\pi^+\pi^-\pi^0$. 
The effective Lagrangians for $J/\psi\rightarrow\rho(1450,1700)\pi^0\rightarrow K^+ K^-\pi^0$ are constructed 
\begin{eqnarray} 
{\cal L}_{J/\psi\rightarrow\rho\pi}& =& {2g_J\over f_\pi}\epsilon^{\mu\nu\alpha\beta}(\partial_\mu 
J_\nu-\partial_\nu J\mu)(\partial_\alpha\rho^i_\beta-\partial_\beta\rho^i\alpha)\pi^i,\nonumber \\
{\cal L}_{\rho\rightarrow K^+ K^-}& = & g_\rho \rho^i_\mu f_{iab} K_a\partial_\mu K_b,
\end{eqnarray}
where $\rho$ is either $\rho(1450)$ or $\rho(1700)$. The distribution of $J/\psi\rightarrow\rho(1450,1700)\pi^0
\rightarrow K^+ K^-\pi^0$ is obtained
\begin{eqnarray}
\lefteqn{\frac{d\Gamma}{dq^2}=\frac{1}{9f^2_\pi m^4_J}\frac{1}{(2\pi)^3}(q^2-4m^2_K)^{{3\over2}}
\{(m^2_J+q^2-m^2_{\pi^0})^2
-4m^2_J q^2\}^{{3\over2}}}\nonumber \\
&&\sum_i|\frac{g_1}{q^2-m^2_1+i\sqrt{q^2}\Gamma_1}+
\frac{g_2}{q^2-m^2_2+i\sqrt{q^2}\Gamma_2}|^2,
\end{eqnarray}
where q is the momentum of $\rho$, \(m_1=1.45GeV\) and \(m_2=1.7GeV\), $g_{1,2}$ are two parameters. 
\(\Gamma_1=0.4GeV\), \(\Gamma_2=0.4GeV\), and \(g_1=1.495 g_2\) are tried to fit 
the distribution obtained by BESII[1]. The width of the structure obtained by Eq.(5) is about 600MeV 
which is compatible with the data(1). In the chiral limit the effective Lagrangian of 
$\rho(1450,1700)\rightarrow\pi\pi$
is constructed as
\begin{equation}
{\cal L}_{\rho(1450,1700)\rightarrow\pi\pi} =  g_\rho \rho^i_\mu \epsilon_{ijk} \pi_j\partial_\mu\pi_k.
\end{equation}
Notice $f_{3ab}={1\over2}(a,b=4,5,6,7)$.
Substituting $m^2_K$ by $m^2_{\pi^+}$ in Eq.(5) and multiplying Eq.(5) by a factor of 4, the distribution
of $J/\psi\rightarrow\rho^0(1450, 1700)\pi^0\rightarrow\pi^+\pi^-\pi^0$
is obtained in the resonance area of $\rho^0(1450,1700)$. Integrating this distribution, it is obtained 
\begin{equation}
BR(J/\psi\rightarrow\rho^0(1450,1700)\pi^0\rightarrow\pi^+\pi^-\pi^0)\sim 8 
BR(J/\psi\rightarrow\rho^0(1450,1700)\pi^0\rightarrow K^+K^-\pi^0).
\end{equation}
On the other hand, we can search for $\rho^{\pm}(1450,1700)$ in $J/\psi\rightarrow\pi^+\pi^-\pi^0$.

The decay mode $\rho(1450,1700)\rightarrow4\pi$ has been found[7]. $\rho\pi\pi$ is the dominant decay
channel of $\rho(1700)$. According to the chiral meson theory[11], 
$a_1$ strongly couples to $\rho\pi$. Therefore, $\rho(1700)\rightarrow\rho\pi\pi$ is dominant by
$\rho(1700)\rightarrow a_1\pi$.  Because of small phase space of
$\rho(1450)\rightarrow a_1(1260)\pi$ $BR(\rho(1450)\rightarrow a_1(1260)\pi\rightarrow\rho\pi\pi)$ is small.
$\rho(1700)$ can be searched in 
$J/\psi\rightarrow\rho(1700)\pi\rightarrow a_1\pi\pi$. 

$\rho(1450,1700)\rightarrow\eta\rho, \omega\pi$ are discovered[7]. $\rho(1450,1700)$
can be found in $J/\psi\rightarrow\rho(1450,1700)\pi\rightarrow\eta\rho\pi, \omega\pi\pi$.  
$\rho(1450,1700)\rightarrow\phi\pi$ are OZI suppressed. Therefore, $BR(\rho(1450,1700)\rightarrow\phi\pi)$
are much smaller than $BR(\rho(1450,1700)\rightarrow\pi\pi, a_1\pi, \eta\pi, \omega\pi)$. However, 
if the broad structure is a four quark state X[3,4] which via "fall apart" decays, there is no suppression
for this four quark state to decay to $\phi\pi$. Larger $BR(J/\psi\rightarrow X\pi\rightarrow\phi\pi\pi)$ 
and very small $BR(J/\psi\rightarrow X\pi\rightarrow\pi\pi\pi, a_1\pi\pi, \eta\rho\pi, \omega\pi\pi)$ should 
be expected if X is a four quark state.

The nature of $\rho(1450,1700)$ is very interesting. In Ref.[15] the authors claim that $\rho(1450)$ has 
a mass consistent with radial 2s, but its decays show characteristics of hybrids, and suggest that 
this state may be a 2s-hybrid mixture. In Ref.[16] it is argued that the inclusion of a isovector hybrid is 
essential to explain the $e^+ e^-\rightarrow 4\pi$ data. 

In summary, the possibility that the broad structure reported by BESII is caused by $\rho(1450,1700)$ 
is not ruled out. Various tests are proposed.


\begin{thebibliography}{40}
\bibitem{} M.Ablikim et al., BES Collaboration, Phys.Rev.Lett.{\bf 97}, 142002(2006).
\bibitem{} F.K.Guo and P.N.Shen, Phys.Rev.{\bf D74},097503(2006).
\bibitem{} M.Karliner and H.Lipkin, hep-ph/0607093; 
           Z.G.Wang and S.L.Wan, hep-ph/0608263; A.Zhang, T.Huang and T.Steele, 
           hep-ph/0612146.
\bibitem{} G.J.Ding and M.L.Yan, Phys.Lett {\bf B643}, 33(2006).
\bibitem{} X.Liu et al., hep-ph/0701022.
\bibitem{} S.Eidelman and J.J.Hernandez-Rey, J.of Physics G, {\bf 33},1(2006).
\bibitem{} Particle Data Group, J.of Physics G, {\bf 33},1(2006).
\bibitem{} A.Abele et al., Crystal Barrel Collaboration, Phys.Lett.{\bf B468},178(1999).
\bibitem{} A.Abele et al., Crystal Barrel Collaboration, Phys.Lett.{\bf B450},275(1999).
\bibitem{} B.A.Li, Phys.Rev.{\bf D52},5165(1995), {\bf 52},5184(1995).
\bibitem{} J.Gao and B.A.Li, Phys.Rev.{\bf D61},113006(2000).
\bibitem{} B.A.Li, Phys.Rev.{\bf D55}, 1436(1997).
\bibitem{} B.A.Li,D.N.Gao and M.L.Yan, Phys.Rev.{\bf D58},09031(1998).
\bibitem{} K.Abe et al., Belle collaboration, hep-ex/0512071.
\bibitem{} T.Barnes et al., Phys.Rev.{\bf D55}, 4157(1997); F.E.Close and P.Page, Phys.Rev.{\bf D56},1584(1997).
\bibitem{} A.Donnachie and A.A.Kalashnikova, Phys.Rev.{\bf D60},114011(1999)
\end{thebibliography}
\end{document}